\documentclass{article}

\usepackage[preprint]{neurips_2026} 

\usepackage[utf8]{inputenc} 
\usepackage[T1]{fontenc}    
\usepackage{hyperref}       
\usepackage{url}            
\usepackage{booktabs}       
\usepackage{amsfonts}       
\usepackage{amsmath}        
\usepackage{amssymb}        
\usepackage{nicefrac}       
\usepackage{microtype}      
\usepackage{xcolor}         
\usepackage{graphicx}       

\title{Verify Before You Fix: Agentic Execution Grounding for Trustworthy Cross-Language Code Analysis}

\author{%
  Jugal Gajjar\thanks{Alternate Email: 812jugalgajjar@gmail.com} \\
  Department of Computer Science\\
  The George Washington University\\
  Washington, DC 20052 \\
  \texttt{jugal.gajjar@gwu.edu} \\
}

\begin{document}

\maketitle

\begin{abstract}
  Learned classifiers deployed in agentic pipelines face a fundamental reliability problem: predictions are probabilistic inferences, not verified conclusions, and acting on them without grounding in observable evidence leads to compounding failures across downstream stages. Software vulnerability analysis makes this cost concrete and measurable. We address this through a unified cross-language vulnerability lifecycle framework built around three LLM-driven reasoning stages---hybrid structural-semantic detection, execution-grounded agentic validation, and validation-aware iterative repair---governed by a strict invariant: no repair action is taken without execution-based confirmation of exploitability. Cross-language generalization is achieved via a Universal Abstract Syntax Tree (uAST) normalizing Java, Python, and C++ into a shared structural schema, combined with a hybrid fusion of GraphSAGE and Qwen2.5-Coder-1.5B embeddings through learned two-way gating, whose per-sample weights provide intrinsic explainability at no additional cost. The framework achieves 89.84--92.02\% intra-language detection accuracy and 74.43--80.12\% zero-shot cross-language F1, resolving 69.74\% of vulnerabilities end-to-end at a 12.27\% total failure rate. Ablations establish necessity: removing uAST degrades cross-language F1 by 23.42\%, while disabling validation increases unnecessary repairs by 131.7\%. These results demonstrate that execution-grounded closed-loop reasoning is a principled and practically deployable mechanism for trustworthy LLM-driven agentic AI.
\end{abstract}

\section{Introduction}
\label{sec:intro}

Large language models and learned classifiers have demonstrated remarkable capability across code understanding, medical diagnosis, legal reasoning, and financial decision-making, yet their deployment in high-stakes settings exposes a fundamental limitation: predictions are probabilistic inferences over training distributions, not verified conclusions about the world. Acting on such predictions without grounding them in observable evidence leads to compounding failures: false positives accumulate, downstream agents expend effort on non-issues, and human operators lose trust through repeated false alarms \cite{marcilio2019static, chustecki2024benefits}. How AI systems should handle uncertainty in consequential multi-stage decisions and communicate the basis for those decisions to human collaborators remains a central open problem in trustworthy and explainable AI.

Software vulnerability analysis offers a uniquely rigorous testbed for these questions. Unlike many domains where ground truth is subjective or delayed, exploitability is verifiable through program execution---providing a concrete feedback signal that AI reasoning systems can use to ground their predictions. The task is inherently multi-stage: a detection model identifies candidates, a reasoning agent confirms genuine exploitability, and a generative model produces a correct fix. Each stage demands a different reasoning modality, structural pattern recognition, hypothesis-driven execution reasoning, and constrained code generation, making it an ideal setting for studying how LLM-driven agents can reason reliably across a decision lifecycle \cite{wang2025vulagent, li2025mavul}. Yet existing approaches fragment this lifecycle into isolated tools: detectors predict without confirming, repair systems act on unverified findings \cite{yang2025survey}, and all major approaches require separate models per programming language \cite{harzevili2023survey, uddin2025deep}.

We present a unified cross-language vulnerability lifecycle framework built around three LLM-driven reasoning stages: hybrid structural-semantic detection, execution-grounded agentic validation, and validation-aware iterative repair. The governing principle is that no repair action is taken without execution-based confirmation---converting probabilistic predictions into evidence-backed decisions. We argue this constitutes a general principle for trustworthy agentic AI: \emph{execution grounding as a mechanism for uncertainty reduction in multi-stage LLM pipelines}. Cross-language generalization is achieved through a Universal Abstract Syntax Tree (uAST) \cite{gajjar2025mlcpd} normalizing Java, Python, and C++ into a shared schema, combined with a hybrid fusion of graph embeddings \cite{hamilton2017inductive} and Qwen2.5-Coder-1.5B \cite{hui2024qwen2} semantic representations via learned two-way gating \cite{gajjar2025bridging}. The per-sample gating weights are directly interpretable, revealing whether structural or semantic reasoning drove each prediction, providing intrinsic explainability at no additional inference cost.

Our contributions are fourfold:

\begin{enumerate}

	\item \textbf{Universal structural representation for cross-language LLM reasoning.} A uAST schema enabling zero-shot transfer across Java, Python, and C++ at 74.43--80.1\% F1, a 23.42\% improvement over language-specific representations without per-language retraining.
	
	\item \textbf{Interpretable hybrid reasoning with modality-aware gating.} A two-way gating fusion of structural and LLM semantic reasoning, achieving 89.84--92.02\% detection accuracy with per-sample explainability at no additional cost.
	
	\item \textbf{Execution-grounded agentic validation as a trustworthy AI mechanism.} An LLM-driven agent that converts classifier predictions into execution-verified decisions. Ablations show removing this stage increases unnecessary repairs by 131.7\% and reduces end-to-end success by 9.56 percentage points.
	
	\item \textbf{Closed-loop iterative repair with human-AI collaboration support.} A validation-aware LLM repair loop acting exclusively on confirmed findings, with structured diagnostic traces surfaced to human reviewers for non-convergent cases.
	
\end{enumerate}

Together, these contributions demonstrate that execution-grounded feedback loops are a principled and empirically effective mechanism for reducing overconfidence in LLM-driven agentic pipelines---establishing a blueprint for trustworthy, explainable, and human-collaborative AI reasoning in high-stakes domains. The code, data splits, and model artifacts will be made available upon request.

\section{Related Work}
\label{sec:related_work}

\subsection{Uncertainty and Evidence Grounding in Agentic AI}

Probabilistic predictions in multi-stage pipelines accumulate errors when downstream agents act without external verification---a failure mode documented across LLM planning \cite{valmeekam2023planbench}, self-refinement \cite{huang2023large}, and tool-augmented reasoning \cite{yao2022react, shinn2023reflexion, gou2023critic}. Verification-before-action has consequently emerged as a core principle in reliable agentic design, with grounded feedback from execution, proof-checkers, or tool outputs measurably improving decision quality over model-internal reasoning alone \cite{gou2023critic}. Software vulnerability analysis instantiates this problem concretely: exploitability is directly verifiable through execution \cite{ullah2025cve}, making it a uniquely rigorous testbed for evidence-grounded agentic AI.

\subsection{LLMs for Code Understanding and Security Analysis}

Early transformer-based code models---CodeBERT \cite{feng2020codebert}, GraphCodeBERT \cite{guo2020graphcodebert}, CodeT5 \cite{wang2021codet5}---established transfer learning for security tasks, while more recent autoregressive models including Code Llama \cite{roziere2023code}, DeepSeek-Coder \cite{guo2024deepseek}, and Qwen2.5-Coder \cite{hui2024qwen2} extend this with richer semantic reasoning over API usage and control-flow intent \cite{zhou2025large}. However, LLMs in security-critical contexts remain susceptible to hallucination \cite{ji2023survey, zhang2025llm} and long-range context limitations \cite{liu2024lost}, motivating hybrid architectures that complement semantic reasoning with structural program representations.

\subsection{Reasoning-Centric and Agentic AI Systems}

Multi-stage reasoning pipelines substantially outperform single-pass prediction in program analysis. VulAgent \cite{wang2025vulagent} and MAVUL \cite{li2025mavul} demonstrate that hypothesis generation and iterative refinement reduce false positives, while agentic CVE reproduction frameworks \cite{ullah2025cve} show that plan-execute-verify reasoning is essential for reliable conclusions. In repair, SAN2PATCH \cite{kim2025logs}, AutoPatch \cite{seo2025autopatch}, and SecureFixAgent \cite{gajjar2025securefixagent} employ structured multi-step reasoning---yet all rely on static analysis for validation rather than execution-grounded evidence, leaving the exploitability confirmation gap unaddressed.

\subsection{Structural Reasoning, Explainability, and Trustworthy AI}

GNN-based detectors---Devign \cite{zhou2019devign}, ReVeal \cite{chakraborty2021deep}, VulGraB \cite{wang2023vulgrab}---introduced structured program reasoning but remain language-specific \cite{harzevili2023survey, uddin2025deep}. Hybrid structural-semantic fusion \cite{gajjar2025bridging} shows complementary gains, yet prior explainability approaches rely on post-hoc attribution \cite{fu2022linevul, arp2022and} rather than intrinsic architectural transparency. Trustworthy deployment of learned classifiers requires moving beyond statistical predictions toward observable, grounded evidence \cite{arp2022and}, a gap that self-refinement approaches \cite{wang2025vulagent, li2025mavul} address only partially, as corrections remain within the model's own reasoning space rather than grounded in execution.

\subsection{Automated Repair and Human-AI Collaboration}

APR has evolved from template-based methods \cite{le2011genprog, long2016automatic} through neural models \cite{chen2019sequencer, lutellier2020coconut} to LLM-driven patch generation \cite{yang2025survey, bhandari2025generating}. A persistent limitation across all paradigms is operating on unverified detector outputs \cite{yang2025survey}---repair effort is allocated without execution-based confirmation of exploitability, and non-convergent cases fail silently without supporting human oversight.

\subsection{Our Positioning}

Our framework enforces execution grounding as a governing invariant across all three lifecycle stages---inference, grounding, and action---unlike prior pipelines that act on unverified predictions \cite{valmeekam2023planbench, huang2023large}. Unlike prior detection approaches, we achieve cross-language transfer through uAST normalization \cite{gajjar2025mlcpd} without per-language retraining. Unlike prior agentic systems, our validation agent grounds predictions in observable execution behavior rather than model-internal reasoning. Unlike prior repair systems, remediation is gated on execution-confirmed exploitability and non-convergent cases are delegated to human reviewers with structured diagnostic traces, making human-AI collaboration a first-class design principle rather than an afterthought.

\begin{table}

  \caption{Comparison of representative approaches across five dimensions. \checkmark\ full support, $\circ$ partial, \texttimes\ absent. Ours is the only approach achieving all five simultaneously; execution grounding is absent from all prior work.}
  \label{tab:related-work-comparison}
  \centering
  \small
  
  \begin{tabular}{lccccc}
  
    \toprule
    Approach & \begin{tabular}[c]{@{}c@{}}Multi-stage\\Lifecycle\end{tabular} & \begin{tabular}[c]{@{}c@{}}Cross-lang.\\Transfer\end{tabular} & \begin{tabular}[c]{@{}c@{}}Execution\\Grounding\end{tabular} & \begin{tabular}[c]{@{}c@{}}Intrinsic\\XAI\end{tabular} & \begin{tabular}[c]{@{}c@{}}Human--AI\\Collab.\end{tabular} \\
    \midrule
    Devign \cite{zhou2019devign}           & \texttimes & \texttimes & \texttimes & \texttimes & \texttimes \\
    GraphCodeBERT \cite{guo2020graphcodebert}   & \texttimes & $\circ$    & \texttimes & \texttimes & \texttimes \\
    VulGraB \cite{wang2023vulgrab}         & \texttimes & \texttimes & \texttimes & \texttimes & \texttimes \\
    LineVul \cite{fu2022linevul}         & \texttimes & \texttimes & \texttimes & $\circ$    & \texttimes \\
    VulAgent \cite{wang2025vulagent}        & \checkmark    & \texttimes & \texttimes & \texttimes & \texttimes \\
    MAVUL \cite{li2025mavul}           & $\circ$    & \texttimes & \texttimes & \texttimes & $\circ$ \\
    AutoPatch \cite{seo2025autopatch}       & $\circ$    & \texttimes & \texttimes & \texttimes & \texttimes \\
    SecureFixAgent \cite{gajjar2025securefixagent}  & $\circ$    & \texttimes & \texttimes & \texttimes & \texttimes \\
    MalCodeAI \cite{gajjar2025malcodeai}       & $\circ$    & \checkmark    & \texttimes & \texttimes & \texttimes \\
    \midrule
    \textbf{Ours}        & \checkmark & \checkmark & \checkmark & \checkmark & \checkmark \\
    \bottomrule
    
  \end{tabular}
  
\end{table}

\section{Proposed Framework}
\label{sec:framework}

\subsection{Overview}

The framework instantiates a general principle for trustworthy agentic AI: a three-stage closed loop of inference, grounding, and action, where each stage gates the next on verified evidence rather than probabilistic prediction. This design is domain-agnostic---the lifecycle applies wherever predictions are verifiable and acting on unconfirmed outputs carries measurable cost.

\begin{figure}

  \centering
  \includegraphics[width=\linewidth]{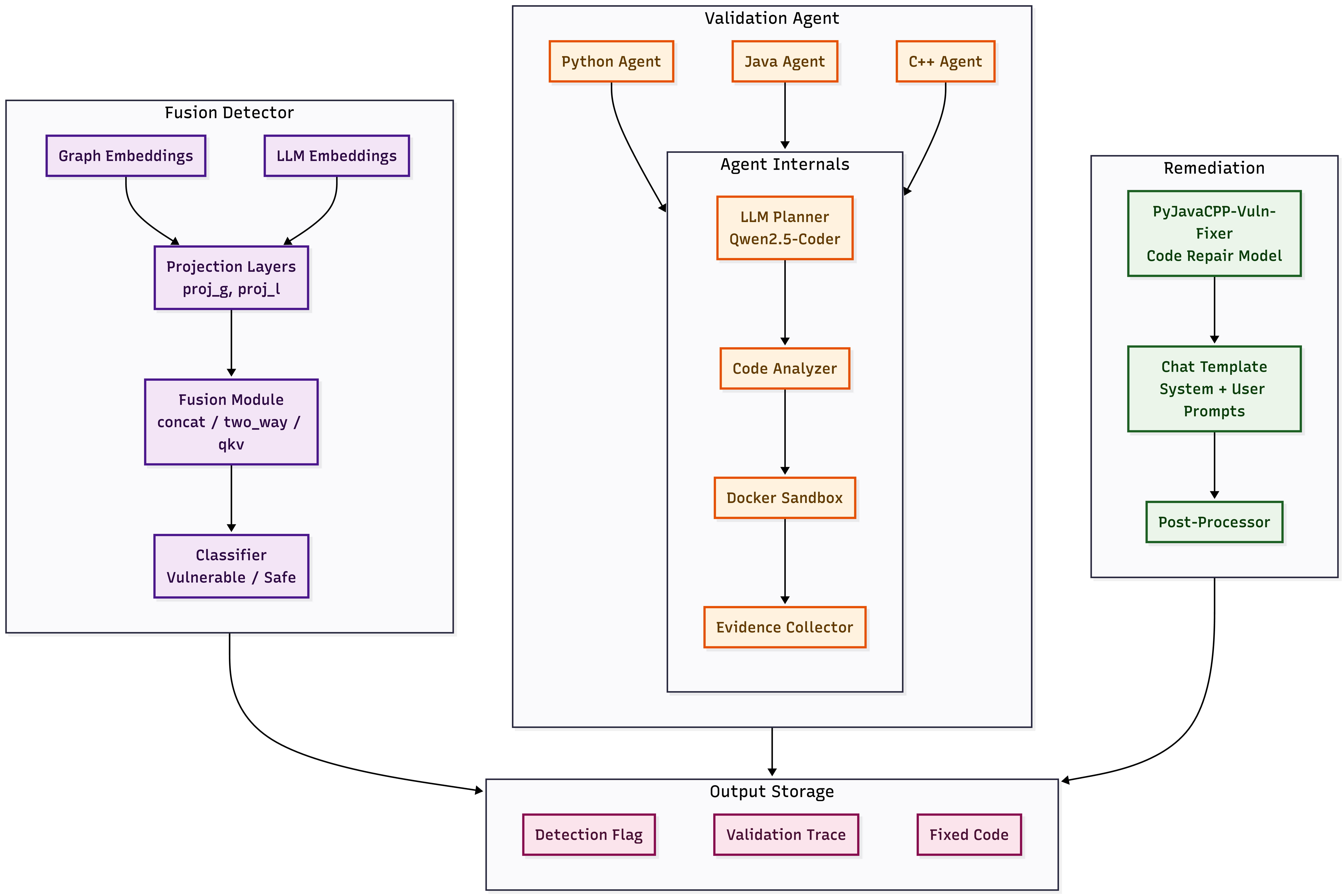}
  \caption{Three-stage lifecycle architecture. The Fusion Detector combines graph and LLM embeddings via two-way gating for binary detection. The Validation Agent confirms exploitability through sandboxed execution. The Remediation module generates minimal patches, outputting a detection flag, validation trace, and fixed code.}
  \label{fig:architecture}
  
\end{figure}

Figure~\ref{fig:architecture} illustrates the three-stage lifecycle framework for our vulnerability analysis testbed. Given a source file, the detection stage parses it into a Universal Abstract Syntax Tree (uAST) and produces a binary vulnerability flag via hybrid fusion of graph and LLM embeddings. Flagged samples enter the validation stage, where language-specific agents---dispatched through a shared LLM planner, Docker sandbox, and evidence collector---confirm genuine exploitability through sandboxed execution. Confirmed samples proceed to iterative repair, where a fine-tuned LLM generates minimal patches under re-detection supervision, with structured diagnostic traces surfaced to human reviewers for non-convergent cases. The governing invariant is strict: no repair action is taken without execution-based confirmation. The following subsections detail the internals of each stage and how they operate within this lifecycle.

\subsection{Universal AST and Hybrid Detection}

Cross-language vulnerability detection faces a fundamental representation mismatch: identical vulnerability patterns manifest through radically different surface syntax across languages. We address this through a Universal Abstract Syntax Tree (uAST) that normalizes Java, Python, and C++ into a shared four-layer schema---metadata, flat node array, universal node taxonomy (47 categories over 200+ language-specific types), and cross-language semantic mapping---constructed via Tree-sitter \cite{brunsfeld2026treesitter} and serialized in Parquet format \cite{gajjar2025mlcpd}.

The hybrid detection model fuses two complementary reasoning branches. The structural branch encodes each uAST as a directed graph processed by GraphSAGE \cite{hamilton2017inductive}, sampling fixed-size neighborhoods through mean aggregation across two message-passing layers to produce a 128-dimensional graph-level embedding. The semantic branch encodes raw source text through Qwen2.5-Coder-1.5B \cite{hui2024qwen2}, averaging the final hidden states across sequence positions to produce a 1536-dimensional embedding. Both embeddings are L2-normalized and projected to a shared 128-dimensional latent space:

\begin{equation}
    \hat{g}_i = \text{LayerNorm}\!\left(\phi\!\left(W_g \tilde{h}_{G,i} + b_g\right)\right), 
    \quad 
    \hat{l}_i = \text{LayerNorm}\!\left(\phi\!\left(W_l \tilde{h}_{L,i} + b_l\right)\right)
\end{equation}

A two-way gating mechanism computes per-sample modality scores via learned scoring networks, normalized through softmax to weights $[\alpha_{g,i},\, \alpha_{l,i}]$ summing to one. The fused representation $\hat{h}_i = \alpha_{g,i}\hat{g}_i + \alpha_{l,i}\hat{l}_i$ drives a binary classifier. These weights are directly interpretable: $\alpha_g \approx 1$ for structurally explicit vulnerabilities such as buffer overflows, $\alpha_l \approx 1$ for logic vulnerabilities requiring semantic context---providing per-prediction explainability intrinsic to the architecture. The model trains with composite loss:

\begin{equation}
    \mathcal{L} = \mathcal{L}_{\text{CE}} 
                + \lambda_{\text{nce}}\,\mathcal{L}_{\text{InfoNCE}} 
                + \lambda_{\text{lap}}\,\mathcal{L}_{\text{Laplacian}}
\end{equation}

\noindent where $\mathcal{L}_{\text{InfoNCE}}$ \cite{oord2018representation} encourages modality alignment and $\mathcal{L}_{\text{Laplacian}}$ \cite{ando2006learning} smooths structural 
embeddings over connected graph regions.

\subsection{Execution-Grounded Agentic Validation}

The validation agent (Figures~\ref{fig:python_agent}--\ref{fig:cpp_agent}) implements a plan-execute-verify reasoning loop that converts probabilistic detector outputs into binary, execution-backed verdicts. Given a flagged sample, an LLM planner (Qwen2.5-Coder-1.5B-Instruct) generates a structured exploit hypothesis specifying attack vector, concrete payloads, and environmental preconditions. A language-specific harness generator synthesizes instrumented test programs---integrating AddressSanitizer and UndefinedBehaviorSanitizer for C++, AspectJ \cite{kiczales2001overview} bytecode instrumentation for Java, and module-level monkeypatching for Python. Harnesses execute within Docker containers under strict resource isolation (memory limits, network isolation, system call monitoring, 30--45 second timeouts).

The agent operates asymmetrically based on detection flags. For flag~$= 1$, active confirmation deploys 3--5 payloads per hypothesis with a 60-second timeout. For flag~$= 0$, limited probing applies 1--2 generic payloads with a 15-second timeout, recovering false negatives while minimizing overhead. Early stopping triggers immediately upon confirmed exploitation, with iterative refinement applied when initial tests are inconclusive---up to a maximum of five attempts. Evidence classification distinguishes confirming, suggestive, and neutral observations; at least one confirming item is required to declare exploitation success.

\subsection{Validation-Aware Iterative Repair}

Repair activates exclusively for execution-confirmed samples. A LoRA fine-tuned \cite{hu2022lora} Qwen2.5-Coder-1.5B-Instruct generates minimal patches from structured prompts containing the vulnerable code, confirmed vulnerability type, successful exploit payload, and observed malicious behavior. Patch application uses AST manipulation libraries---redbaron \cite{redbaron2026} for Python, javaparser \cite{javaparser2026} for Java, libclang \cite{libclang2026} for C++---ensuring syntactic validity and localized scope. Re-detection parses the patched code into uAST and re-evaluates through the full hybrid model; flag~$= 0$ terminates the loop with success, flag~$= 1$ triggers the next iteration with accumulated context from prior attempts.

The loop runs for a maximum of five iterations. Successfully repaired samples are verified through differential analysis confirming that only vulnerability-implicated regions were modified. Non-convergent samples, those exhausting the iteration budget, are flagged for human review with structured diagnostic traces including all attempted patches, rejection reasons, and persistent vulnerability indicators, positioning the system as a human-AI collaborative tool at the boundary of autonomous capability.

\section{Experiments and Results}
\label{sec:experiments_results}

\subsection{Experimental Setup}

Experiments are conducted across Java, Python, and C++ on a balanced dataset of 120,000 code files, 60,000 vulnerable (paired with fixes) and 60,000 safe, distributed equally at 40,000 samples per language, split 80/10/10 for train/validation/test. The dataset integrates real-world vulnerability databases (PrimeVul \cite{ding2024vulnerability}, MegaVul \cite{ni2024megavul}, DiverseVul \cite{chen2023diversevul}, CVEfixes \cite{bhandari2021cvefixes}, ReposVul \cite{wang2024reposvul}, Vul4J \cite{bui2022vul4j}, PySecDB \cite{sun2023exploring}, among others; see Appendix~\ref{appendix:data_details}) with \textasciitilde20\% synthetic samples generated by GPT-4.1 and Claude Sonnet 4.5 to ensure balanced vulnerability type coverage across SQL injection, command injection, path traversal, and insecure deserialization. The selected detection architecture---GraphSAGE \cite{hamilton2017inductive} paired with Qwen2.5-Coder-1.5B \cite{hui2024qwen2}---was chosen through systematic evaluation of nine combinations (3 graph encoders $\times$ 3 LLMs), achieving the highest hybrid F1 of 90.21\% (Appendix~\ref{appendix:arch_selection}). All experiments run on Google Colab Pro with NVIDIA A100/H100 GPUs; local inference uses an Apple M4 MacBook Pro (48GB) via the MLX framework.

\subsection{Detection Results}

\begin{table}[h]

  \caption{Zero-shot cross-language detection results. The hybrid model outperforms both baselines across all six pairs, with C++~$\rightarrow$~Java achieving the highest F1 (0.7818). Structural-only transfer degrades toward near-random, confirming uAST normalization as essential for cross-language generalization.}
  \label{tab:cross-language}
  \centering
  \small
  
  \begin{tabular}{lcccccc}
  
    \toprule
    & \multicolumn{2}{c}{\textbf{Structural-Only}} & \multicolumn{2}{c}{\textbf{Semantic-Only}} & \multicolumn{2}{c}{\textbf{Hybrid}} \\
    \cmidrule(r){2-3} \cmidrule(r){4-5} \cmidrule(r){6-7}
    \textbf{Train $\rightarrow$ Test} & \textbf{F1} & \textbf{Acc} & \textbf{F1} & \textbf{Acc} & \textbf{F1} & \textbf{Acc} \\
    \midrule
    Java $\rightarrow$ Python   & 0.5521 & 59.34\% & 0.6824 & 71.86\% & 0.7443 & 76.84\% \\
    Java $\rightarrow$ C++      & 0.5287 & 57.12\% & 0.7016 & 73.42\% & 0.7692 & 78.97\% \\
    Python $\rightarrow$ Java   & 0.5734 & 61.03\% & 0.6892 & 72.51\% & 0.7525 & 77.63\% \\
    Python $\rightarrow$ C++    & 0.5469 & 58.41\% & 0.6984 & 72.94\% & 0.7716 & 79.12\% \\
    C++ $\rightarrow$ Java      & 0.5196 & 56.48\% & 0.7129 & 74.08\% & 0.7818 & 80.12\% \\
    C++ $\rightarrow$ Python    & 0.5348 & 57.36\% & 0.7041 & 73.27\% & 0.7593 & 78.45\% \\
    \midrule
    Average    & 0.5426 & 58.29\% & 0.6981 & 73.01\% & 0.7631 & 78.52\% \\
    \bottomrule
    
  \end{tabular}
  
\end{table}

\begin{figure}[t]

  \centering
  \includegraphics[width=\linewidth]{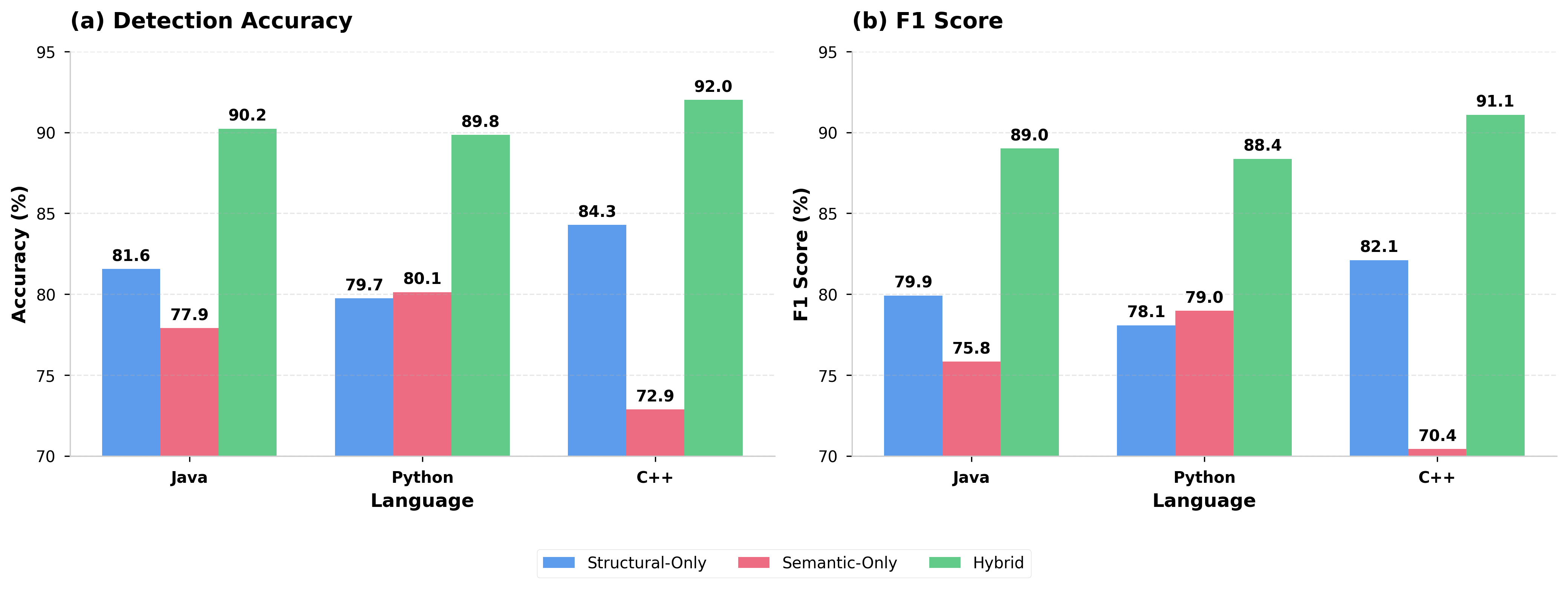}
  \caption{Intra-language detection performance. The hybrid model achieves 89.84--92.02\% accuracy and 0.8837--0.9109 F1, outperforming both baselines by 8--15 percentage points. C++ achieves the highest accuracy (92.02\%), reflecting explicit structural patterns in memory management vulnerabilities.}
  \label{fig:intra-language}
  
\end{figure}

The hybrid model achieves 89.84--92.02\% intra-language accuracy and 0.8837--0.9109 F1 across all three languages (Figure~\ref{fig:intra-language}), outperforming structural-only and semantic-only baselines by 8--15 percentage points. Cross-language transfer (Table~\ref{tab:cross-language}) demonstrates meaningful zero-shot generalization at 74.43--80.12\% F1---a 23.42\% average improvement over language-specific representations. C++ achieves the highest intra-language accuracy (92.02\%), reflecting explicit structural patterns in memory management vulnerabilities. Semantic-only models exhibit poor cross-language transfer, confirming that uAST structural normalization is the primary driver of generalization.

\subsection{Validation and Repair}

\begin{table}[h]
  
  \caption{Validation effectiveness across languages. For flag~$= 1$, the agent confirms exploitability in 66.84--71.49\% of cases while rejecting 58.72--62.37\% of false positives. For flag~$= 0$, limited probing recovers 11.76--16.21\% of missed vulnerabilities at a spurious rate of only 2.67--3.98\%.}
  \label{tab:validation}
  \centering
  \small
  
  \begin{tabular}{lcccc}
  
    \toprule
    & \multicolumn{2}{c}{\textbf{Flag $= 1$}} & \multicolumn{2}{c}{\textbf{Flag $= 0$}} \\
    \cmidrule(r){2-3} \cmidrule(r){4-5}
    \textbf{Language} & \textbf{Exploit Confirm} & \textbf{FP Rejection} & \textbf{FN Recovery} & \textbf{Spurious Rate} \\
    \midrule
    Java    & 70.28\% & 60.43\% & 13.87\% & 2.94\% \\
    Python  & 71.49\% & 62.37\% & 16.21\% & 2.67\% \\
    C++     & 66.84\% & 58.72\% & 11.76\% & 3.98\% \\
    \midrule
    Average & 69.54\% & 60.51\% & 13.95\% & 3.20\% \\
    \bottomrule
    
  \end{tabular}
  
\end{table}

\begin{table}[h]

	\caption{Repair performance across languages. Python achieves the highest success rate (87.27\%, 2.3 iterations) while C++ is most challenging (81.37\%, 3.4 iterations). Post-repair pass rates of 90.44--93.15\% confirm patches robustly eliminate vulnerabilities.}
	\label{tab:repair}
	\centering
	\small
    
	\begin{tabular}{lcccc}
    
		\toprule
		\textbf{Language} & \textbf{Repair Success} & \textbf{Avg Iterations} & \textbf{Max-Iter Failures} & \textbf{Post-Repair Pass} \\
		\midrule
		Java     & 85.14\% & 2.7 & 10.38\% & 91.36\% \\
		Python   & 87.27\% & 2.3 & \ 8.17\% & 93.15\% \\
		C++      & 81.37\% & 3.4 & 14.79\% & 90.44\% \\
		\midrule
		Combined & 83.78\% & 2.8 & 11.12\% & 91.63\% \\
		\bottomrule
        
	\end{tabular}
    
\end{table}

Validation confirms 66.84--71.49\% of flagged samples as genuinely exploitable while rejecting 58.72--62.37\% of detector false positives (Table~\ref{tab:validation}), preventing unnecessary repair effort on non-exploitable findings. Python achieves the highest confirmation rate through dynamic typing simplifying automated test construction; C++ exhibits the lowest, reflecting the complexity of triggering memory-level exploits. Repair succeeds in 81.37--87.27\% of cases within the five-iteration budget (Table~\ref{tab:repair}), converging in 2.3--3.4 iterations on average. Post-repair detection pass rates of 90.44--93.15\% confirm patches robustly eliminate vulnerabilities. Language difficulty follows Python (easiest), Java (moderate), C++ (hardest)---consistent with manual repair complexity.

\subsection{End-to-End Performance and Ablation}

\begin{table}[h]

  \caption{End-to-end results and ablation study. Full pipeline resolves 69.74\% of vulnerabilities at 12.27\% failure rate. Removing uAST degrades cross-language F1 by 23.42\%; disabling validation increases unnecessary repairs by 131.7\% and reduces end-to-end success by 9.56 pp---establishing both components as essential.}
  \label{tab:e2e-ablation}
  \centering
  \small
  \setlength{\tabcolsep}{10pt}
  \renewcommand{\arraystretch}{1.25}
  
  \begin{tabular}{llc}
  
    \toprule
    \multicolumn{1}{c}{\textbf{Configuration}} & \multicolumn{1}{c}{\textbf{Metric}} & \textbf{Value} \\
    \midrule
    \multicolumn{3}{l}{\textit{Full system (end-to-end pipeline)}} \\[2pt]
    & Resolved vulnerabilities    & 69.74\% \\
    & False positives eliminated  & 61.24\% \\
    & Unnecessary repairs avoided & 73.13\% \\
    & Total pipeline failure      & 12.27\% \\
    \midrule
    \multicolumn{3}{l}{\textit{Ablation A: remove uAST normalization}} \\[2pt]
    & Avg intra-language F1 drop  & $-$3.79\%  \\
    & Avg cross-language F1 drop  & $-$23.42\% \\
    \midrule
    \multicolumn{3}{l}{\textit{Ablation B: disable execution-grounded validation}} \\[2pt]
    & Unnecessary repairs (avg)   & $+$131.7\% \\
    & End-to-end success drop     & $-$9.56 pp \\
    \bottomrule
    
  \end{tabular}
  
\end{table}

\begin{figure}[t]

  \centering
  \includegraphics[width=0.9\linewidth]{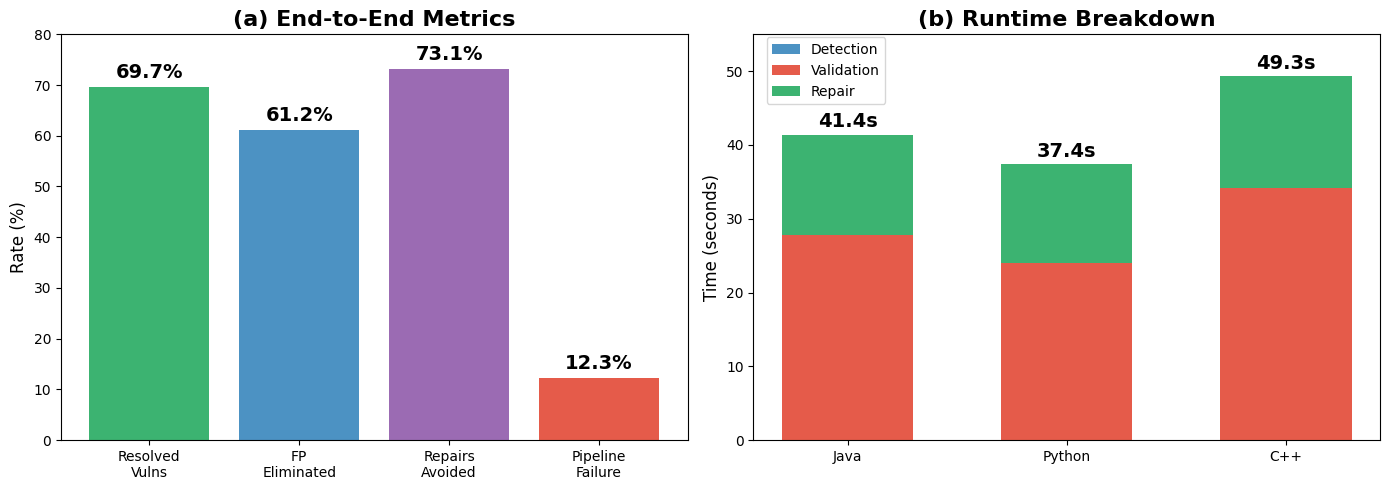}
  \caption{End-to-end pipeline metrics. The full system resolves 69.74\% of vulnerabilities, eliminates 61.24\% of false positives, avoids 73.13\% of unnecessary repairs, and maintains a 12.27\% total failure rate---demonstrating stable multi-stage integration.}
  \label{fig:e2e}
  
\end{figure}

The full pipeline resolves 69.74\% of vulnerabilities end-to-end at a 12.27\% total failure rate (Figure~\ref{fig:e2e}), with validation eliminating 61.24\% of detector false positives before remediation and avoiding 73.13\% of unnecessary repairs. The 69.74\% end-to-end resolution reflects compounding multi-stage error rates---each stage operating at high but imperfect accuracy---rather than any single point of failure. Ablation studies (Table~\ref{tab:e2e-ablation}) validate the necessity of both core components. Removing uAST normalization degrades average cross-language F1 by 23.42\% while incurring only 3.79\% intra-language loss, confirming it as the primary driver of cross-language generalization. Disabling execution-grounded validation increases unnecessary repairs by 131.7\% and reduces end-to-end success by 9.56 percentage points, directly quantifying the cost of prediction-only pipelines in high-stakes agentic settings. The full system processes 1,700--2,400 samples daily on a single A100 GPU (37--49 seconds per sample), enabling practical CI/CD integration without cloud dependencies.

\section{Discussion}
\label{sec:discussion}

\subsection{Execution Grounding as a General Trustworthy AI Principle}

Our ablation results, +131.7\% unnecessary repairs and $-$9.56 percentage points end-to-end success without validation, quantify a cost that extends beyond vulnerability analysis: \emph{agentic LLM pipelines acting on probabilistic predictions without external verification systematically accumulate errors across stages}. This mirrors failure modes documented in tool-augmented agents \cite{wang2025vulagent, li2025mavul} and self-refinement systems, where corrections remain within the model's own reasoning space rather than grounded in observable outcomes. Our plan-execute-verify loop demonstrates that introducing a single execution-grounded verification stage between prediction and action measurably improves reliability---a principle transferable to any agentic pipeline operating in domains with verifiable ground truth, including theorem proving, scientific hypothesis testing, and code generation with test-based feedback \cite{gou2023critic}. More broadly, any domain where a binary oracle exists---unit tests, formal checkers, simulators, or experimental outcomes---can instantiate this loop with domain-specific grounding mechanisms in place of execution.

\subsection{Intrinsic Explainability through Architectural Fusion}

The per-sample gating weights ($\alpha_g, \alpha_l$) produced by our two-way fusion mechanism reveal which reasoning modality---structural or semantic---drove each detection decision, without any post-hoc attribution overhead. This contrasts with the dominant XAI paradigm of applying gradient-based or attention-based explanation methods after prediction \cite{chakraborty2021deep, fu2022linevul}, which are known to be unstable under input perturbations \cite{arp2022and}. Our results show that $\alpha_g \approx 1$ for structurally explicit vulnerabilities and $\alpha_l \approx 1$ for semantic logic flaws, a consistent, faithful signal that arises naturally from the architecture. This suggests that designing fusion mechanisms with interpretable weighting is a practical path toward intrinsic XAI in any multi-modal reasoning system---from medical imaging fused with clinical text to code structure fused with semantics---more faithful than post-hoc methods applied to black-box models.

\subsection{Structural Abstraction for Cross-Domain Transfer in Learned Systems}

The 23.42\% cross-language F1 improvement from uAST normalization, with only 3.79\% intra-language cost, illustrates a general representation learning principle: task-relevant structural abstraction enables transfer that purely semantic representations cannot achieve alone. Semantic-only models fail at cross-language transfer because they encode surface lexical patterns that do not generalize across syntactic conventions. The uAST addresses this by lifting language-specific constructs into a shared structural schema, analogous to how intermediate representations enable transfer in other domains, compiler IRs across hardware targets \cite{lattner2004llvm}, universal dependency parsing across natural languages, and modality-agnostic embeddings in multimodal learning. Learned systems operating across heterogeneous input domains consistently benefit from such abstraction layers, and our results provide concrete empirical grounding for this principle in the code reasoning setting.

\subsection{Limitations and Future Work}

Three limitations bound the current framework---and reflect broader challenges in execution-grounded agentic systems generally: bounded reasoning horizon in multi-hop inference chains (2-hop GraphSAGE insufficient for complex taint flows), incomplete environmental reconstruction in sandboxed verification (producing false negatives on complex deserialization and format string vulnerabilities), and coverage gaps in evaluation (concurrency, cryptographic misuse, and business logic flaws remain unevaluated). Looking forward, training the validation agent via reinforcement learning with exploit confirmation as a reward signal could improve discovery rates for complex patterns. Expanded language coverage is immediately feasible as Tree-sitter provides grammars for 40+ languages and the uAST schema is already mapped for ten \cite{gajjar2025mlcpd}. Integrating formal verification for patch correctness would elevate repair from vulnerability removal to provably safe patch synthesis, closing the remaining gap between autonomous AI capability and production deployment. Future research in execution-grounded agentic AI should prioritize addressing these limitations to extend such frameworks beyond security toward broader high-stakes domains.

\section{Conclusion}
\label{sec:conclusion}

We presented a unified cross-language vulnerability lifecycle framework demonstrating that execution-grounded feedback loops are a principled and empirically effective mechanism for reducing overconfidence in LLM-driven agentic pipelines. By integrating hybrid structural-semantic detection, autonomous execution-grounded validation, and validation-aware iterative repair into a closed-loop system, we achieve 89.84--92.02\% intra-language detection accuracy, 74.43--80.12\% zero-shot cross-language F1, and 69.74\% end-to-end vulnerability resolution with ablations directly quantifying the cost of removing each component. Beyond software security, the core architectural principles established here---execution grounding as uncertainty reduction, structural abstraction for cross-domain transfer, and intrinsic explainability through fusion design---are broadly applicable to any agentic AI system operating in high-stakes domains where predictions must be verified before action is taken. As LLM-driven agents are increasingly deployed in consequential multi-stage decision pipelines, building in mechanisms that know when to verify, when to repair, and when to defer to human judgment becomes not an engineering detail but a fundamental requirement for trustworthy AI.

\bibliographystyle{plain}
\bibliography{references}


\newpage
\appendix

\section{Dataset Details}
\label{appendix:data_details}

Table~\ref{tab:dataset-sources} provides the complete dataset composition. The full dataset integrates 15 real-world vulnerability databases and repositories supplemented with synthetic samples. Approximately 80\% of samples (96,000) are sourced from real-world repositories and vulnerability databases; the remaining 20\% (24,000) are synthetically generated using GPT-4.1 and Claude Sonnet 4.5 to ensure balanced representation across vulnerability types and coding patterns. All samples are distributed equally across Java, Python, and C++ at 40,000 samples per language (20,000 vulnerable + 20,000 safe).

\begin{table}[h]
    \caption{Complete dataset sources by language coverage. \checkmark\ denotes inclusion. The dataset integrates 15 real-world sources across curated vulnerability databases, repository-mined commits, and synthetic generation, ensuring comprehensive coverage of SQL injection, command injection, path traversal, and insecure deserialization across all three languages.}
    \label{tab:dataset-sources}
    \centering
    \small
    \setlength{\tabcolsep}{8pt}
    \renewcommand{\arraystretch}{1.2}
    \begin{tabular}{lccc}
        \toprule
        \textbf{Source} & \textbf{Java} & \textbf{Python} & \textbf{C++} \\
        \midrule
        PrimeVul \cite{ding2024vulnerability}                          & \checkmark &            &            \\
        MegaVul \cite{ni2024megavul}                             & \checkmark &            &            \\
        DiverseVul \cite{chen2023diversevul}                      &            &            & \checkmark \\
        Vul4J \cite{bui2022vul4j}                                & \checkmark &            &            \\
        PySecDB \cite{sun2023exploring}                             &            & \checkmark &            \\
        ReposVul \cite{wang2024reposvul}                          & \checkmark & \checkmark & \checkmark \\
        CVEfixes \cite{bhandari2021cvefixes}                      & \checkmark & \checkmark & \checkmark \\
        Source Code Vulnerability \cite{saratov2026scv}          &            & \checkmark & \checkmark \\
        Vulnerable Programming Dataset \cite{thakur2026vpd}      &            & \checkmark & \checkmark \\
        Vulnerability Fix Dataset \cite{biswas2026vfd}           & \checkmark &            &            \\
        cmonplz/Python\_Vulnerability\_Remediation \cite{cmonplz2026pvr}    &            & \checkmark &            \\
        Shyyshawarma/cppvul \cite{shyyshawarma2026cppvul}                      &            &            & \checkmark \\
        MarioVar/vulnerable-code\_chitchat\_doss1232 \cite{mariovar2026chitchat}     &            &            & \checkmark \\
        CyberNative/Code\_Vulnerability\_Security\_DPO \cite{cybernative2026dpo}  & \checkmark & \checkmark &            \\
        lemon42-ai/Code\_Vulnerability\_Labeled\_Dataset \cite{oumida2025labeled}     & \checkmark & \checkmark & \checkmark \\
        \midrule
        GPT-4.1 synthetic                                         & \checkmark & \checkmark & \checkmark \\
        Claude Sonnet 4.5 synthetic                               & \checkmark & \checkmark & \checkmark \\
        \bottomrule
    \end{tabular}
\end{table}

\section{Universal AST Schema}
\label{appendix:uast_schema}

The uAST organizes each source file into four hierarchical layers designed around core properties like \textit{losslessness} (preserving every syntactic element without semantic compression), \textit{uniformity} (enforcing a consistent JSON schema across all languages), and \textit{queryability} (enabling direct node addressing and category-based indexing without recursive traversal).

The \textit{metadata layer} captures global file characteristics, including language, node count, and a SHA-256 content hash for deduplication. The \textit{flat node array} linearizes the tree into $O(1)$-addressable nodes, each storing type, text content, source span, parent index, and child indices. The \textit{universal node taxonomy} abstracts 200+ language-specific types into 47 universal categories---\texttt{FUNCTION\_DECLARATION}, \texttt{VARIABLE\_ASSIGNMENT}, \texttt{CONTROL\_FLOW}, and others---enabling Python's \texttt{def}, Java's \texttt{public static void}, and C++'s function definitions to be queried uniformly. The \textit{cross-language semantic mapping layer} translates language-specific constructs (Python's \texttt{with}, Java's try-with-resources, C++'s RAII) into shared universal roles while preserving original syntax.

Construction uses Tree-sitter \cite{brunsfeld2026treesitter}, selected over rule-based and regex-based approaches for its complete syntactic coverage, uniform cross-language interface, and robust error recovery. The pipeline comprises six deterministic stages: language detection, recursive AST extraction, node categorization, cross-language mapping, schema validation, and Parquet serialization. Processing throughput averages 600--800 files per minute on multi-core systems.

\section{Implementation Details}
\label{appendix:implementation_details}

The framework is implemented in Python 3.11, totaling approximately 13,000 lines organized into modular components. Core dependencies include PyTorch 2.9.0, Tree-sitter 0.21.0 \cite{brunsfeld2026treesitter}, Transformers 4.57.1, and Docker 29.2.0. The GraphSAGE \cite{hamilton2017inductive} encoder comprises two \texttt{SAGEConv} layers (PyTorch Geometric) with mean aggregation, sampling up to ten neighbors per node. The first layer maps 768-dimensional input features to 256-dimensional hidden representations; the second produces 128-dimensional node embeddings. Global mean pooling aggregates node embeddings into graph-level representations with Xavier uniform initialization and batch normalization after each aggregation step.

Semantic embeddings are extracted from Qwen2.5-Coder-1.5B in fp16 precision with 4,096 token maximum sequence length. For files exceeding this limit, a sliding window approach processes overlapping segments and averages embeddings. The fusion model trains with AdamW (learning rate 10$^{-3}$, weight decay 10$^{-4}$, batch size 64), with early stopping on validation F1 (patience 5 epochs, typically triggering around epoch 25--30). Contrastive loss temperature $\tau = 0.07$; regularization weights $\lambda_{\text{nce}} = 0.1$, $\lambda_{\text{lap}} = 0.01$.

Docker sandbox configurations enforce strict resource isolation: 1--2GB memory limits (language-dependent), 0.9 CPU core quota, \texttt{--network=none} isolation, read-only project mounts, \texttt{--tmpfs /tmp:size=256m}, PID limit of 256, and \texttt{no-new-privileges} security option. Standard streams are captured and truncated to 240KB. The LoRA fine-tuned repair model uses rank 16, adding only 12MB trainable parameters over the base Qwen2.5-Coder-1.5B-Instruct, fine-tuned on 60,000 vulnerability-patch pairs with temperature 0.2 and top-$p$ 0.9 at inference.

\section{Architecture Selection}
\label{appendix:arch_selection}

\begin{table}[h]

    \caption{Hybrid model architecture selection across nine combinations of graph encoder and language model backbone. F1 scores reported on the combined validation set. GraphSAGE paired with Qwen2.5-Coder-1.5B achieves the highest F1 (90.21\%) and is selected for all experiments. GraphSAGE consistently outperforms GCN and GAT across all LLM configurations through inductive neighborhood aggregation suited to irregular node-degree distributions in uAST graphs.}
    \label{tab:arch_selection}
    \centering
    \small
    \setlength{\tabcolsep}{10pt}
    \renewcommand{\arraystretch}{1.25}
    
    \begin{tabular}{lccc}
    
        \toprule
        \textbf{Graph Encoder} & \textbf{Qwen3-0.6B} & \textbf{DeepSeek-Coder-1.3B} & \textbf{Qwen2.5-Coder-1.5B} \\
        \midrule
        GCN \cite{kipf2016semi}           & 0.8524 & 0.8607 & 0.8749 \\
        GraphSAGE \cite{hamilton2017inductive} & 0.8915 & 0.8858 & \textbf{0.9021} \\
        GAT \cite{velivckovic2017graph}      & 0.8673 & 0.8843 & 0.8926 \\
        \bottomrule
        
    \end{tabular}
    
\end{table}

\begin{figure}[h]

  \centering
  \includegraphics[width=\linewidth]{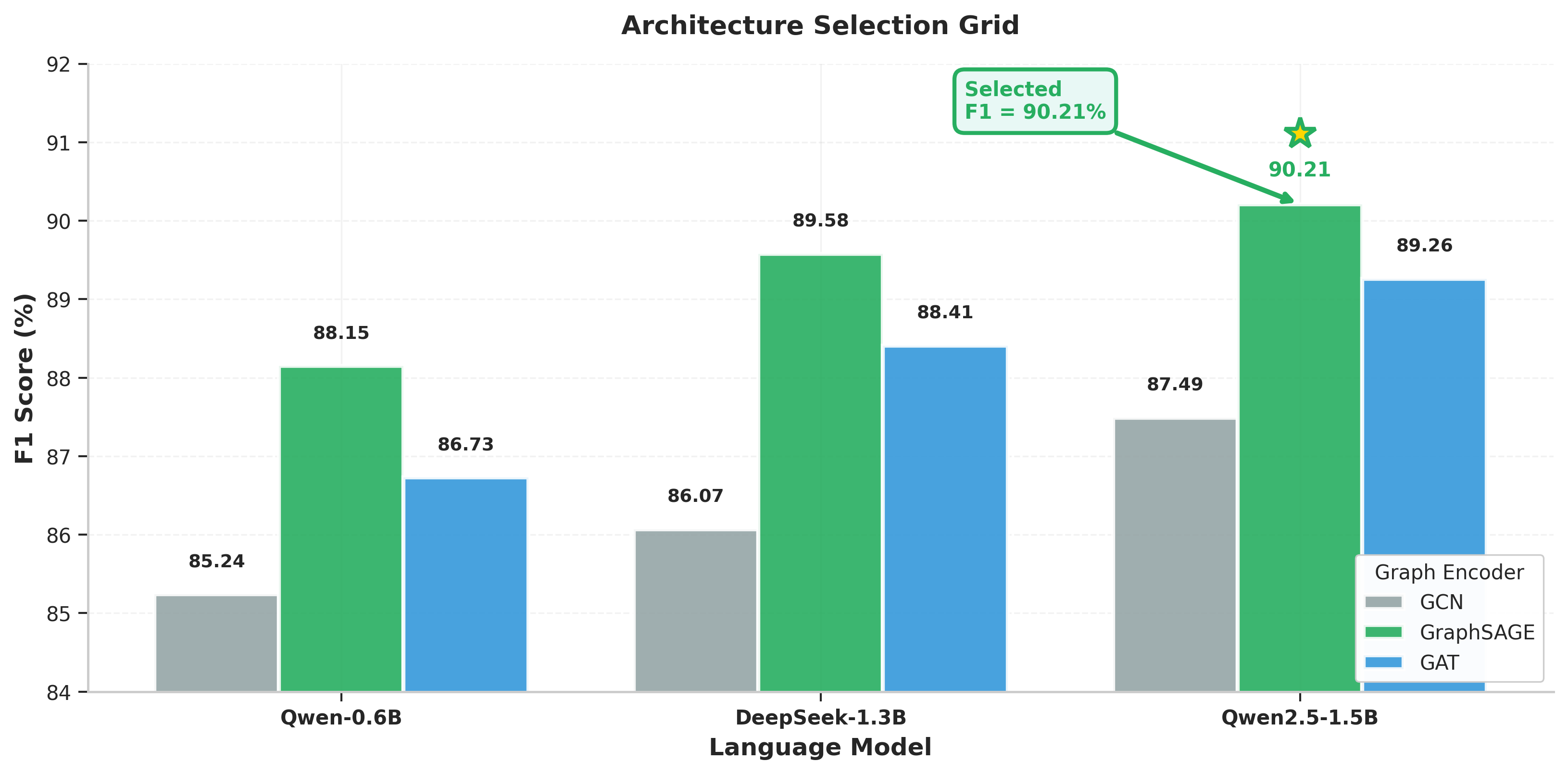}
  \caption{Architecture selection grid comparing graph encoders and LLM backbones across nine combinations. GraphSAGE + Qwen2.5-Coder-1.5B achieves the highest F1 (90.21\%, marked with $\bigstar$) and is selected for all experiments.}
  \label{fig:arch_selection}
  
\end{figure}

\section{Detailed Experimental Results}
\label{appendix:detail_results}

\begin{table}[h]
    \caption{Intra-language detection accuracy and F1 scores. The hybrid model consistently achieves the highest accuracy across all languages, confirming complementary contributions of structural and semantic modalities. C++ achieves the highest hybrid accuracy (92.02\%), reflecting explicit structural patterns in memory management vulnerabilities.}
    \label{tab:intra_language}
    \centering
    \small
    \setlength{\tabcolsep}{8pt}
    \renewcommand{\arraystretch}{1.25}
    \begin{tabular}{lcccccc}
        \toprule
        & \multicolumn{2}{c}{\textbf{Structural-Only}} & \multicolumn{2}{c}{\textbf{Semantic-Only}} & \multicolumn{2}{c}{\textbf{Hybrid}} \\
        \cmidrule(r){2-3} \cmidrule(r){4-5} \cmidrule(r){6-7}
        \textbf{Language} & \textbf{Acc} & \textbf{F1} & \textbf{Acc} & \textbf{F1} & \textbf{Acc} & \textbf{F1} \\
        \midrule
        Java   & 81.56\% & 0.7992 & 77.90\% & 0.7584 & 90.23\% & 0.8901 \\
        Python & 79.74\% & 0.7808 & 80.13\% & 0.7897 & 89.84\% & 0.8837 \\
        C++    & 84.28\% & 0.8211 & 72.87\% & 0.7043 & 92.02\% & 0.9109 \\
        \bottomrule
    \end{tabular}
\end{table}

\begin{table}[h]
    \caption{Per-language repair performance. Python achieves highest success (87.27\%) and fastest convergence (2.3 iterations); C++ is most challenging (81.37\%, 3.4 iterations). Post-repair detection pass rates of 90.44--93.15\% confirm patches robustly eliminate vulnerabilities.}
    \label{tab:repair_detailed}
    \centering
    \small
    \setlength{\tabcolsep}{8pt}
    \renewcommand{\arraystretch}{1.25}
    \begin{tabular}{lcccc}
        \toprule
        \textbf{Language} & \textbf{Repair Success} & \textbf{Avg Iterations} & \textbf{Max-Iter Failures} & \textbf{Post-Repair Pass} \\
        \midrule
        Java     & 85.14\% & 2.7 & 10.38\% & 91.36\% \\
        Python   & 87.27\% & 2.3 &  8.17\% & 93.15\% \\
        C++      & 81.37\% & 3.4 & 14.79\% & 90.44\% \\
        \midrule
        Combined & 83.78\% & 2.8 & 11.12\% & 91.63\% \\
        \bottomrule
    \end{tabular}
\end{table}

\begin{table}[h]
    \caption{Average runtime per sample across pipeline stages. Validation dominates total execution time despite early stopping optimizations, reflecting Docker container startup, compilation, and harness execution overhead. Full pipeline runtime of 37.4--49.3 seconds per sample enables processing approximately 1,700--2,400 samples per day on a single A100 GPU.}
    \label{tab:runtime}
    \centering
    \small
    \setlength{\tabcolsep}{10pt}
    \renewcommand{\arraystretch}{1.25}
    \begin{tabular}{lcccc}
        \toprule
        \textbf{Language} & \textbf{Detection (s)} & \textbf{Validation (s)} & \textbf{Repair (s)} & \textbf{Total (s)} \\
        \midrule
        Java   & 0.18 & 28.6 & 13.6 & 42.38 \\
        Python & 0.16 & 24.8 & 12.4 & 37.36 \\
        C++    & 0.21 & 34.0 & 15.1 & 49.31 \\
        \bottomrule
    \end{tabular}
\end{table}

\begin{figure}[h]
  \centering
  \includegraphics[width=\linewidth]{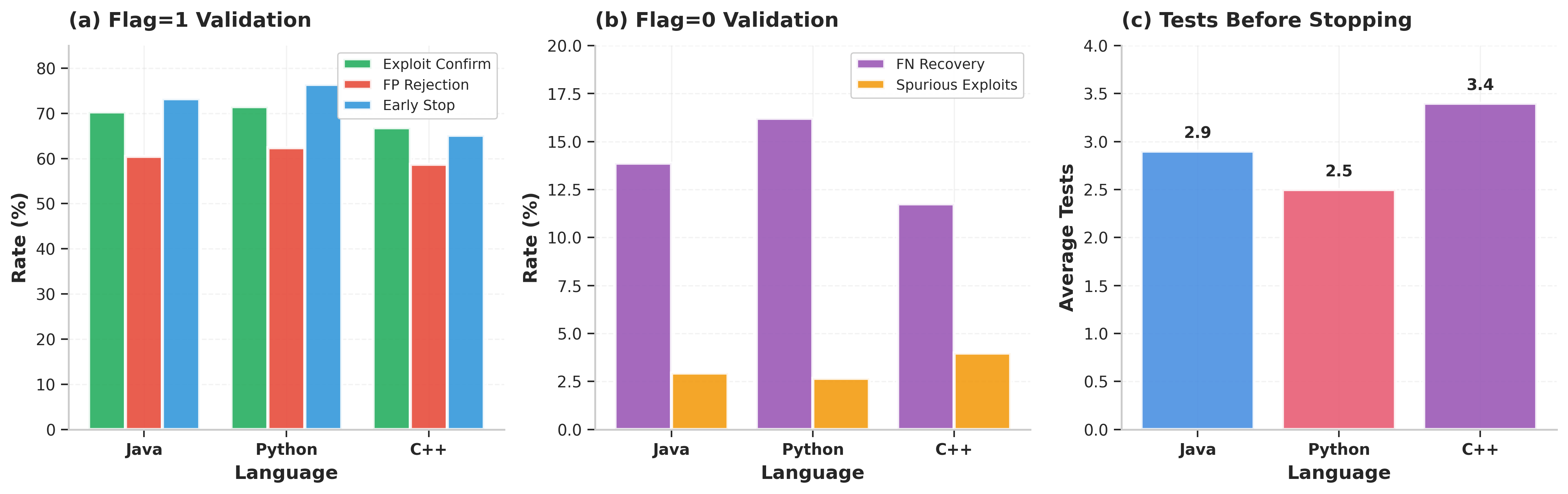}
  \caption{Validation performance across languages. (a) Flag~$= 1$: exploit confirmation rates of 66.84--71.49\%, false positive rejection of 58.72--62.37\%, and early stopping frequencies of 65.13--76.44\%. (b) Flag~$= 0$: missed vulnerability recovery of 11.76--16.21\% at spurious exploit rates of 2.67--3.98\%. (c) Average tests before stopping: 2.5--3.4 across languages.}
  \label{fig:validation}
\end{figure}

\begin{figure}[h]
    \centering
    \includegraphics[width=\linewidth]{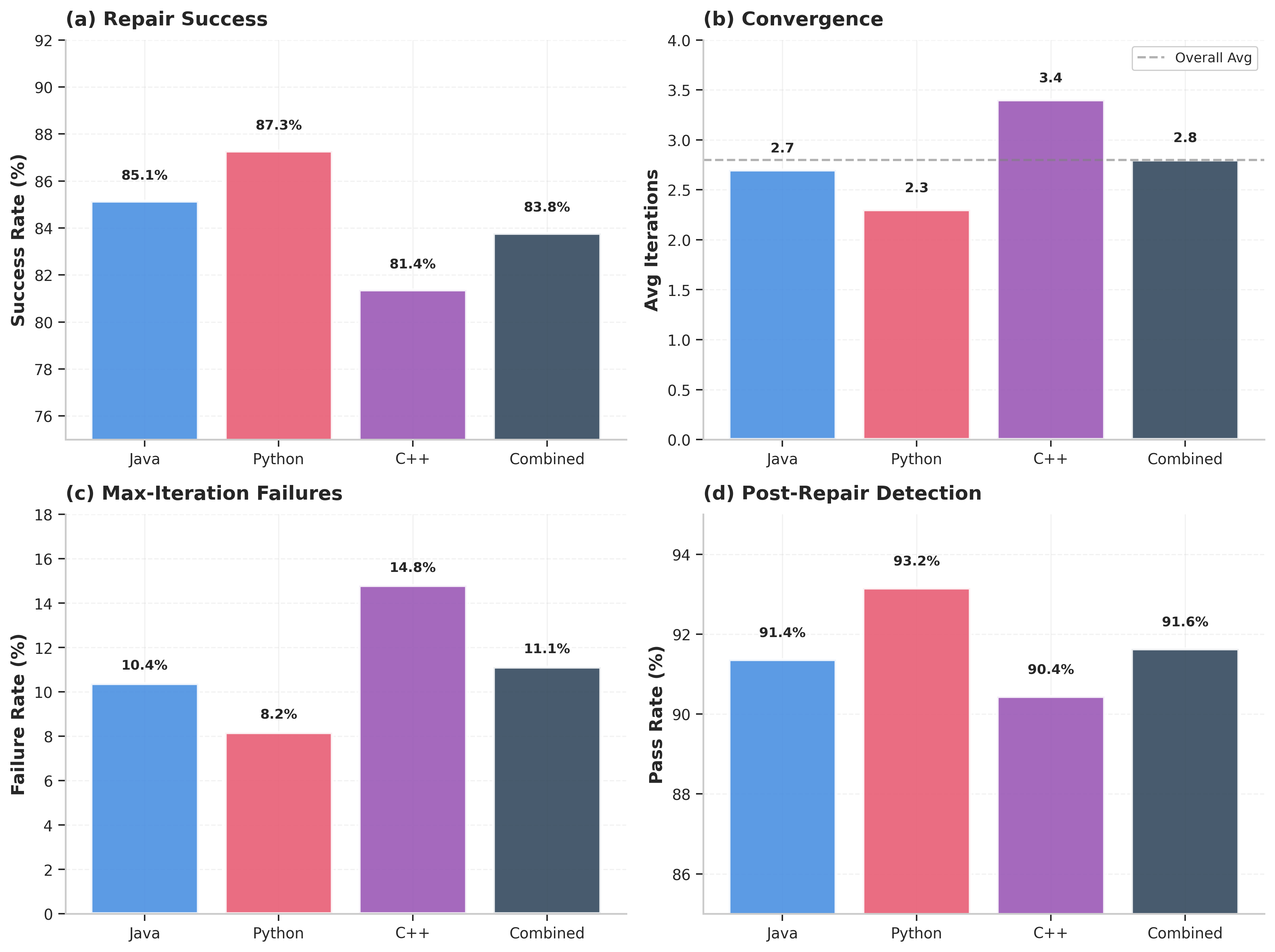}
    \caption{Repair performance metrics across languages. (a) Success rates: 81.37--87.27\%. (b) Average iterations to convergence: 2.3--3.4. (c) Max-iteration failures: 8.17--14.79\%. (d) Post-repair detection pass rates: 90.44--93.15\%.}
    \label{fig:remediation}
\end{figure}

\section{Language-Specific Validation Agent Architectures}
\label{appendix:lang_specific_agent}

\begin{figure}[h]
    \centering
    \includegraphics[width=\linewidth]{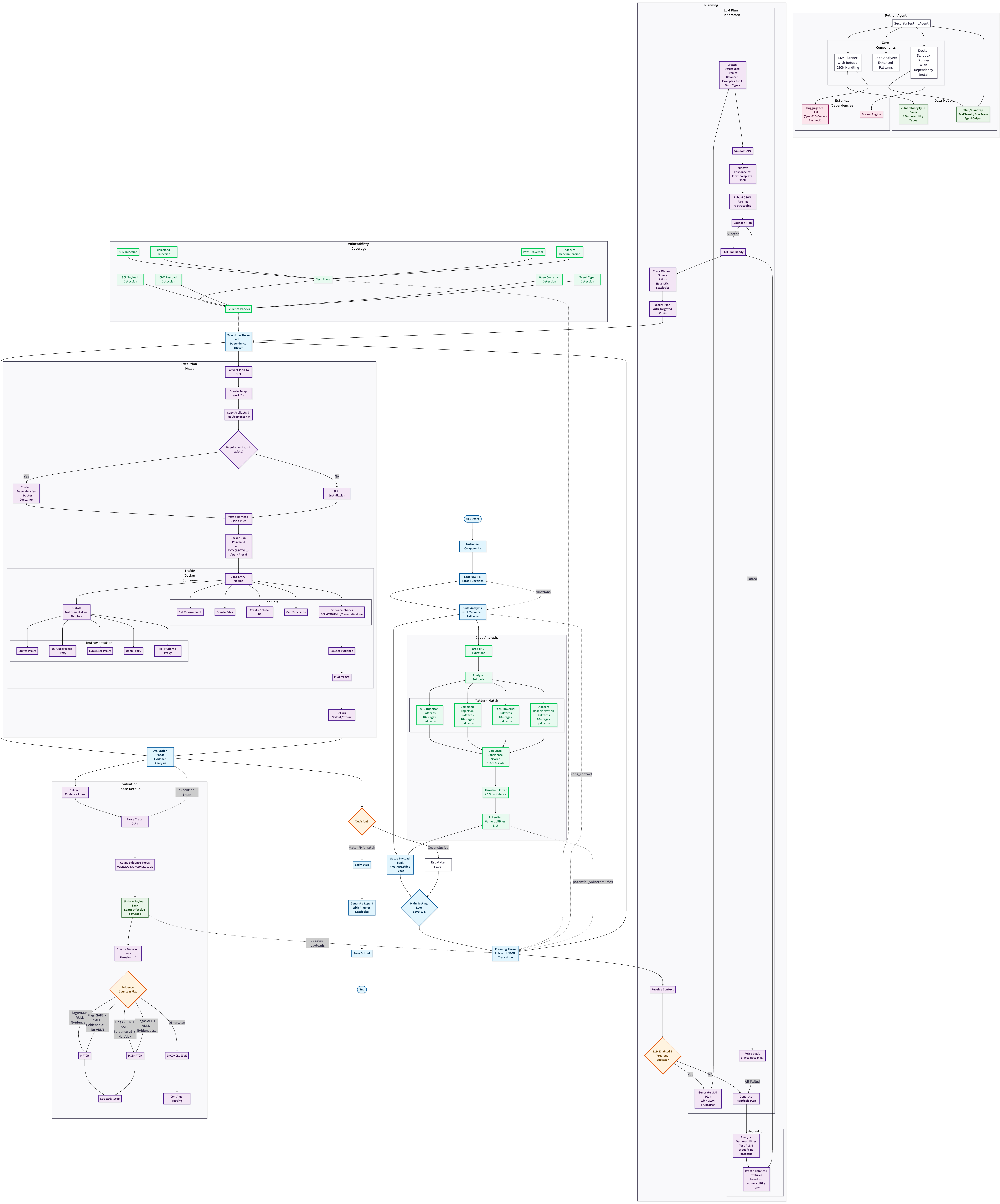}
    \caption{Python validation agent workflow showing hypothesis generation, payload construction, sandboxed execution with runtime monkeypatching instrumentation, and exploit confirmation through behavior observation.}
    \label{fig:python_agent}
\end{figure}

\begin{figure}[h]
    \centering
    \includegraphics[width=\linewidth]{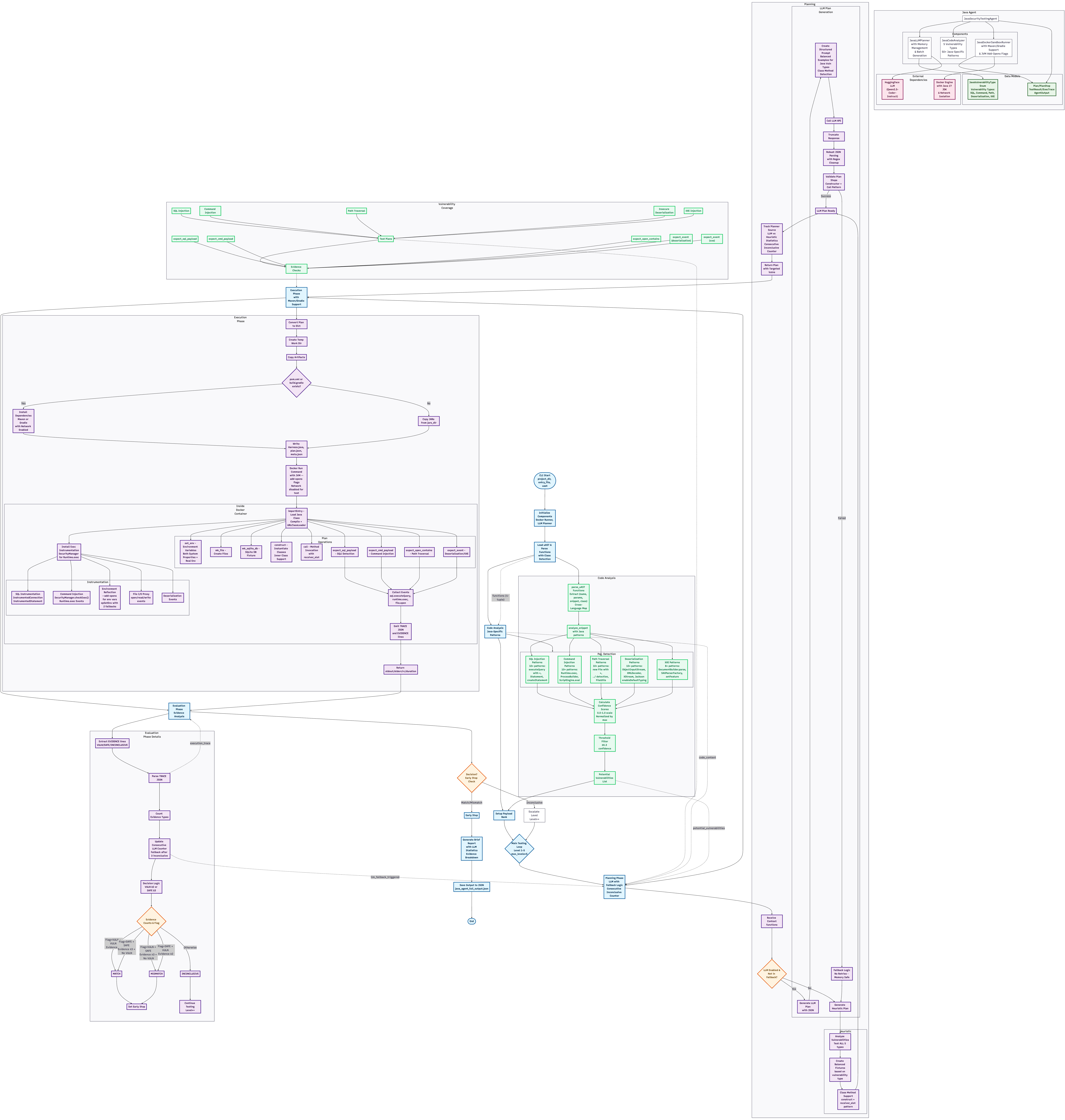}
    \caption{Java validation agent workflow utilizing bytecode instrumentation via AspectJ \cite{kiczales2001overview}, JDBC query logging, Security Manager integration, and heap analysis for tainted data propagation tracking.}
    \label{fig:java_agent}
\end{figure}

\begin{figure}[h]
    \centering
    \includegraphics[width=\linewidth]{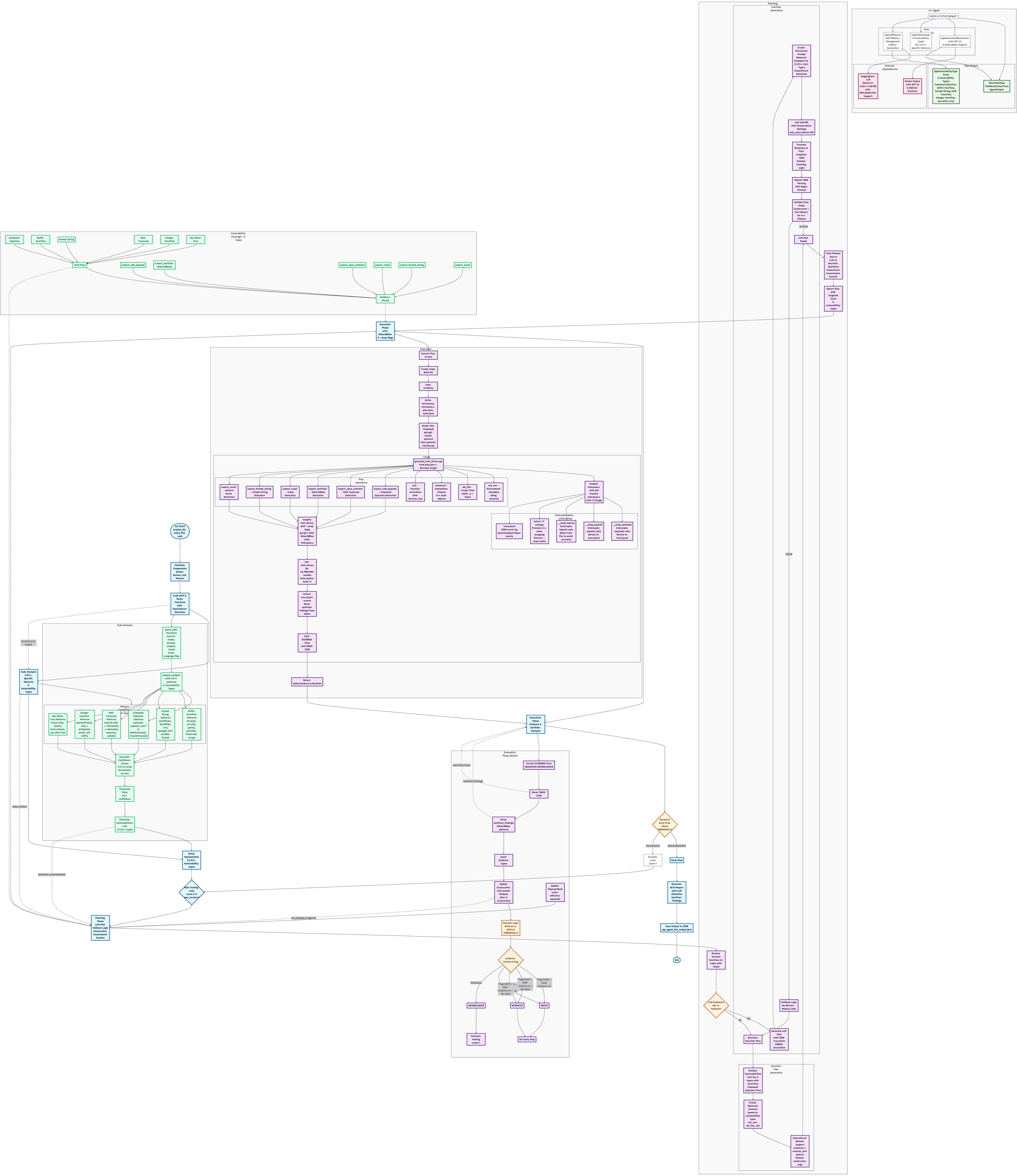}
    \caption{C++ validation agent workflow utilizing AddressSanitizer, UndefinedBehaviorSanitizer, and library call interposition via \texttt{LD\_PRELOAD} for memory safety violation detection and system call monitoring.}
    \label{fig:cpp_agent}
\end{figure}

Figures~\ref{fig:python_agent}, \ref{fig:java_agent}, and \ref{fig:cpp_agent} provide the complete workflow diagrams for each language-specific validation agent. While all three share the same plan-execute-verify loop and evidence classification logic, their instrumentation strategies differ significantly across language runtimes.

The \textbf{Python agent} leverages Python's dynamic nature for comprehensive runtime instrumentation. Database monitoring patches \texttt{sqlite3.Cursor.execute} at module level; file access tracking instruments the built-in \texttt{open} function through monkeypatching; command injection detection wraps \texttt{os.system} and \texttt{subprocess} functions. Exception analysis captures full stack traces with local variable contexts on uncaught exceptions.

The \textbf{Java agent} employs JVM-level instrumentation. JDBC query logging instruments \texttt{PreparedStatement} and \texttt{Statement} classes through AspectJ \cite{kiczales2001overview} bytecode weaving. Reflection monitoring tracks \texttt{Method.invoke}, \texttt{Constructor.newInstance}, and \texttt{Field.set} operations. A restrictive \texttt{SecurityManager} detects unauthorized filesystem operations, network connections, and native library loading. Heap analysis traces tainted data propagation through object references via the JVM Tool Interface.

The \textbf{C++ agent} integrates compiler-level sanitizers. AddressSanitizer (compiled with \texttt{-fsanitize=address}) detects heap/stack/global buffer overflows, use-after-free, and double-free. UndefinedBehaviorSanitizer catches signed integer overflow, null pointer dereference, and misaligned memory access. Library interposition via \texttt{LD\_PRELOAD} replaces security-sensitive functions (\texttt{strcpy}, \texttt{strcat}, \texttt{system}, \texttt{popen}, \texttt{fopen}, \texttt{mysql\_query}) logging all invocations before delegating to original implementations.

\end{document}